\documentclass[aps,prl,twocolumn,showpacs,superscriptaddress,groupedaddress]{revtex4}  
\usepackage{graphicx}  
\usepackage{dcolumn}   
\usepackage{bm}        
\usepackage{amssymb}   

\usepackage{color,soul} 

\begin{document}

\title{Improved $\Lambda p$ Elastic Scattering Cross Sections Between 0.9 and 2.0 GeV/c and Connections to the Neutron Star Equation of State}
                             
\newcommand*{\ANL}{Argonne National Laboratory, Argonne, Illinois 60439}
\newcommand*{\ANLindex}{1}
\affiliation{\ANL}
\newcommand*{\CSUDH}{California State University, Dominguez Hills, Carson, CA 90747}
\newcommand*{\CSUDHindex}{2}
\affiliation{\CSUDH}
\newcommand*{\CANISIUS}{Canisius College, Buffalo, NY}
\newcommand*{\CANISIUSindex}{3}
\affiliation{\CANISIUS}
\newcommand*{\CMU}{Carnegie Mellon University, Pittsburgh, Pennsylvania 15213}
\newcommand*{\CMUindex}{4}
\affiliation{\CMU}
\newcommand*{\CUA}{Catholic University of America, Washington, D.C. 20064}
\newcommand*{\CUAindex}{5}
\affiliation{\CUA}
\newcommand*{\SACLAY}{IRFU, CEA, Universit\'{e} Paris-Saclay, F-91191 Gif-sur-Yvette, France}
\newcommand*{\SACLAYindex}{6}
\affiliation{\SACLAY}
\newcommand*{\CNU}{Christopher Newport University, Newport News, Virginia 23606}
\newcommand*{\CNUindex}{7}
\affiliation{\CNU}
\newcommand*{\UCONN}{University of Connecticut, Storrs, Connecticut 06269}
\newcommand*{\UCONNindex}{8}
\affiliation{\UCONN}
\newcommand*{\DUKE}{Duke University, Durham, North Carolina 27708-0305}
\newcommand*{\DUKEindex}{9}
\affiliation{\DUKE}
\newcommand*{\DUQUESNE}{Duquesne University, 600 Forbes Avenue, Pittsburgh, PA 15282 }
\newcommand*{\DUQUESNEindex}{10}
\affiliation{\DUQUESNE}
\newcommand*{\FU}{Fairfield University, Fairfield CT 06824}
\newcommand*{\FUindex}{11}
\affiliation{\FU}
\newcommand*{\FERRARAU}{Universit\`{a} di Ferrara , 44121 Ferrara, Italy}
\newcommand*{\FERRARAUindex}{12}
\affiliation{\FERRARAU}
\newcommand*{\FIU}{Florida International University, Miami, Florida 33199}
\newcommand*{\FIUindex}{13}
\affiliation{\FIU}
\newcommand*{\FSU}{Florida State University, Tallahassee, Florida 32306}
\newcommand*{\FSUindex}{14}
\affiliation{\FSU}
\newcommand*{\GWUI}{The George Washington University, Washington, DC 20052}
\newcommand*{\GWUIindex}{15}
\affiliation{\GWUI}
\newcommand*{\INFNFE}{INFN, Sezione di Ferrara, 44100 Ferrara, Italy}
\newcommand*{\INFNFEindex}{16}
\affiliation{\INFNFE}
\newcommand*{\INFNFR}{INFN, Laboratori Nazionali di Frascati, 00044 Frascati, Italy}
\newcommand*{\INFNFRindex}{17}
\affiliation{\INFNFR}
\newcommand*{\INFNGE}{INFN, Sezione di Genova, 16146 Genova, Italy}
\newcommand*{\INFNGEindex}{18}
\affiliation{\INFNGE}
\newcommand*{\INFNRO}{INFN, Sezione di Roma Tor Vergata, 00133 Rome, Italy}
\newcommand*{\INFNROindex}{19}
\affiliation{\INFNRO}
\newcommand*{\INFNTUR}{INFN, Sezione di Torino, 10125 Torino, Italy}
\newcommand*{\INFNTURindex}{20}
\affiliation{\INFNTUR}
\newcommand*{\INFNPAV}{INFN, Sezione di Pavia, 27100 Pavia, Italy}
\newcommand*{\INFNPAVindex}{21}
\affiliation{\INFNPAV}
\newcommand*{\ORSAY}{Universit\'{e} Paris-Saclay, CNRS/IN2P3, IJCLab, 91405 Orsay, France}
\newcommand*{\ORSAYindex}{22}
\affiliation{\ORSAY}
\newcommand*{\Juelich}{Institute fur Kernphysik (Juelich), Juelich, Germany}
\newcommand*{\Juelichindex}{23}
\affiliation{\Juelich}
\newcommand*{\JMU}{James Madison University, Harrisonburg, Virginia 22807}
\newcommand*{\JMUindex}{24}
\affiliation{\JMU}
\newcommand*{\KNU}{Kyungpook National University, Daegu 41566, Republic of Korea}
\newcommand*{\KNUindex}{25}
\affiliation{\KNU}
\newcommand*{\LAMAR}{Lamar University, 4400 MLK Blvd, PO Box 10046, Beaumont, Texas 77710}
\newcommand*{\LAMARindex}{26}
\affiliation{\LAMAR}
\newcommand*{\MISS}{Mississippi State University, Mississippi State, MS 39762-5167}
\newcommand*{\MISSindex}{27}
\affiliation{\MISS}
\newcommand*{\ITEP}{National Research Centre Kurchatov Institute - ITEP, Moscow, 117259, Russia}
\newcommand*{\ITEPindex}{28}
\affiliation{\ITEP}
\newcommand*{\UNH}{University of New Hampshire, Durham, New Hampshire 03824-3568}
\newcommand*{\UNHindex}{29}
\affiliation{\UNH}
\newcommand*{\NMSU}{New Mexico State University, PO Box 30001, Las Cruces, NM 88003, USA}
\newcommand*{\NMSUindex}{30}
\affiliation{\NMSU}
\newcommand*{\NSU}{Norfolk State University, Norfolk, Virginia 23504}
\newcommand*{\NSUindex}{31}
\affiliation{\NSU}
\newcommand*{\OHIOU}{Ohio University, Athens, Ohio  45701}
\newcommand*{\OHIOUindex}{32}
\affiliation{\OHIOU}
\newcommand*{\ODU}{Old Dominion University, Norfolk, Virginia 23529}
\newcommand*{\ODUindex}{33}
\affiliation{\ODU}
\newcommand*{\JLU}{II Physikalisches Institut der Universitaet Giessen, 35392 Giessen, Germany}
\newcommand*{\JLUindex}{34}
\affiliation{\JLU}
\newcommand*{\ROMAII}{Universit\`{a} di Roma Tor Vergata, 00133 Rome Italy}
\newcommand*{\ROMAIIindex}{35}
\affiliation{\ROMAII}
\newcommand*{\MSU}{Skobeltsyn Institute of Nuclear Physics, Lomonosov Moscow State University, 119234 Moscow, Russia}
\newcommand*{\MSUindex}{36}
\affiliation{\MSU}
\newcommand*{\SCAROLINA}{University of South Carolina, Columbia, South Carolina 29208}
\newcommand*{\SCAROLINAindex}{37}
\affiliation{\SCAROLINA}
\newcommand*{\TEMPLE}{Temple University,  Philadelphia, PA 19122 }
\newcommand*{\TEMPLEindex}{38}
\affiliation{\TEMPLE}
\newcommand*{\JLAB}{Thomas Jefferson National Accelerator Facility, Newport News, Virginia 23606}
\newcommand*{\JLABindex}{39}
\affiliation{\JLAB}
\newcommand*{\UTFSM}{Universidad T\'{e}cnica Federico Santa Mar\'{i}a, Casilla 110-V Valpara\'{i}so, Chile}
\newcommand*{\UTFSMindex}{40}
\affiliation{\UTFSM}
\newcommand*{\INSUBRIA}{Universit\`{a} degli Studi dell'Insubria, 22100 Como, Italy}
\newcommand*{\INSUBRIAindex}{41}
\affiliation{\INSUBRIA}
\newcommand*{\BRESCIA}{Universit\`{a} degli Studi di Brescia, 25123 Brescia, Italy}
\newcommand*{\BRESCIAindex}{42}
\affiliation{\BRESCIA}
\newcommand*{\GLASGOW}{University of Glasgow, Glasgow G12 8QQ, United Kingdom}
\newcommand*{\GLASGOWindex}{43}
\affiliation{\GLASGOW}
\newcommand*{\YORK}{University of York, York YO10 5DD, United Kingdom}
\newcommand*{\YORKindex}{44}
\affiliation{\YORK}
\newcommand*{\VT}{Virginia Tech, Blacksburg, Virginia   24061-0435}
\newcommand*{\VTindex}{45}
\affiliation{\VT}
\newcommand*{\VIRGINIA}{University of Virginia, Charlottesville, Virginia 22901}
\newcommand*{\VIRGINIAindex}{46}
\affiliation{\VIRGINIA}
\newcommand*{\WM}{College of William and Mary, Williamsburg, Virginia 23187-8795}
\newcommand*{\WMindex}{47}
\affiliation{\WM}
\newcommand*{\YEREVAN}{Yerevan Physics Institute, 375036 Yerevan, Armenia}
\newcommand*{\YEREVANindex}{48}
\affiliation{\YEREVAN}

\newcommand*{\CATANIA}{INFN, Sezione di Catania, 95123 Catania, Italy}
\newcommand*{\CATANIAindex}{49}
\affiliation{\CATANIA}
\newcommand*{\MESSINA}{Universi\`{a} degli Studi di Messina, 98166 Messina, Italy}
\newcommand*{\MESSINAindex}{50}
\affiliation{\MESSINA}

\newcommand*{\MIT}{Massachusetts Institute of Technology, Cambridge, Massachusetts 02139}
\newcommand*{\MITindex}{51}
\affiliation{\MIT}

\newcommand*{\NOWHAMPTON}{Hampton University, Hampton, VA 23668}
\newcommand*{\NOWOHIOU}{Ohio University, Athens, Ohio  45701}
\newcommand*{\NOWISU}{Idaho State University, Pocatello, Idaho 83209}
\newcommand*{\NOWBRESCIA}{Universit\`{a} degli Studi di Brescia, 25123 Brescia, Italy}
\newcommand*{\NOWJLAB}{Thomas Jefferson National Accelerator Facility, Newport News, Virginia 23606}

\author{J.~Rowley}
\affiliation{\OHIOU}
\author{N.~Compton}
\affiliation{\OHIOU}
\author{C.~Djalali}
\affiliation{\OHIOU}
\author{K.~Hicks}
\affiliation{\OHIOU}
\author{J.~Price}
\affiliation{\CSUDH}
\author{N.~Zachariou}
\affiliation{\YORK}

\author {K.P.~Adhikari} 
\affiliation{\ODU}
\author {W.R.~Armstrong} 
\affiliation{\ANL}
\author {H.~Atac} 
\affiliation{\TEMPLE}
\author {L.~Baashen} 
\affiliation{\FIU}
% ...

\collaboration{The CLAS Collaboration}
\noaffiliation
\date{\today}

\begin{abstract}
Strange matter is believed to exist in the cores of neutron stars based on simple kinematics. If this is true, then hyperon-nucleon interactions will play a significant part in the neutron star equation of state (EOS). Yet, compared to other elastic scattering processes, there is very little data on $\Lambda$-$N$ scattering. This experiment utilized the CLAS detector to study the $\Lambda p \rightarrow \Lambda p$ elastic scattering cross section in the incident $\Lambda$ momentum range 0.9-2.0 GeV/c. This is the first data on this reaction in several decades. The new cross sections have significantly better accuracy and precision than the existing world data, and the techniques developed here can also be used in future experiments.
\end{abstract}

\pacs{13.75.Ev Hyperon-nucleon interactions --- 13.85.Lg Total cross section --- 25.55.Cj Elastic and inelastic scattering}
\maketitle

The mass and radius of a neutron star (NS) are important parameters to determine its equation of state (EOS) \cite{Lattimer}.
The presence of strange matter in the core of the NS can have a significant impact on its EOS \cite{Lonardoni}.
One of the biggest challenges with the appearance of hyperons in the stellar core is how to reconcile the LIGO-Virgo results \cite{LIGOVirgo}, which suggest relatively small radii with the existence of massive neutron stars.
However, the presence of hyperons in the core will soften the EOS, and a stiffening of the EOS at the highest densities required to explain massive stars is added by theory without firm experimental justification \cite{Piekarewicz}.
Theoretical models \cite{Haidenbauer} suggest that a combination of $\Lambda N$ and $\Lambda N N$ interactions can create a neutron star consistent with what is observed.
To constrain these observables, better data are needed for $\Lambda$p elastic scattering \cite{Vidana}.

Nucleon-nucleon ($NN$) scattering is perhaps the most well studied of all nuclear reactions.
However, less is known about the scattering of hyperons ($\Lambda$ or $\Sigma$ baryons) from the proton.
Previous data for the scattering of hyperons from the nucleon dates back to the bubble chamber era of the 1960s and 1970s \cite{Crawford,Price}.
These experiments ranged from tens of events to a few hundred spread across multiple momentum bins.
The experiment with the greatest statistics is from \cite{Hauptman} with a total of 584 events spread over 12 momentum bins.
The paucity of data for $\Lambda p$ elastic scattering is due to the difficulty of creating a $\Lambda$ beam.
The decay parameter for $\Lambda$ particles is $c \tau \sim 7.8$~cm, which is too short for any modern beamline.
Hence, data for $\Lambda N$ scattering are very limited in comparison with other elastic scattering processes \cite{PDG}, such as $NN$, $KN$ or $\pi N$.
The present results in this work improve on the existing world data.

The data for this work were collected during the \textit{g12} experiment, which was conducted at the Continuous Electron Beam Accelerator Facility (CEBAF) using the Large Acceptance Spectrometer (CLAS) at the Thomas Jefferson National Accelerator Facility \cite{CLAS}.
The experiment consisted of a photon beam incident on a liquid-hydrogen target.
The photon beam energy ranged from 1.2-5.4 GeV.

The CLAS detector was based on a toroidal magnetic field consisting of six independent sectors separated by the superconducting magnet coils.
Each sector included three regions of drift chambers (DC) to measure the charged  particle trajectories\cite{tagger}.
Plastic scintillators surrounded the DC, which allowed for time-of-flight (TOF) measurements.
The DC and TOF were used to identify the final-state particles and measure their four momenta.
Further detector details can be found in Ref. \cite{g12}.

\begin{figure}[h]
    \includegraphics[width=\linewidth]{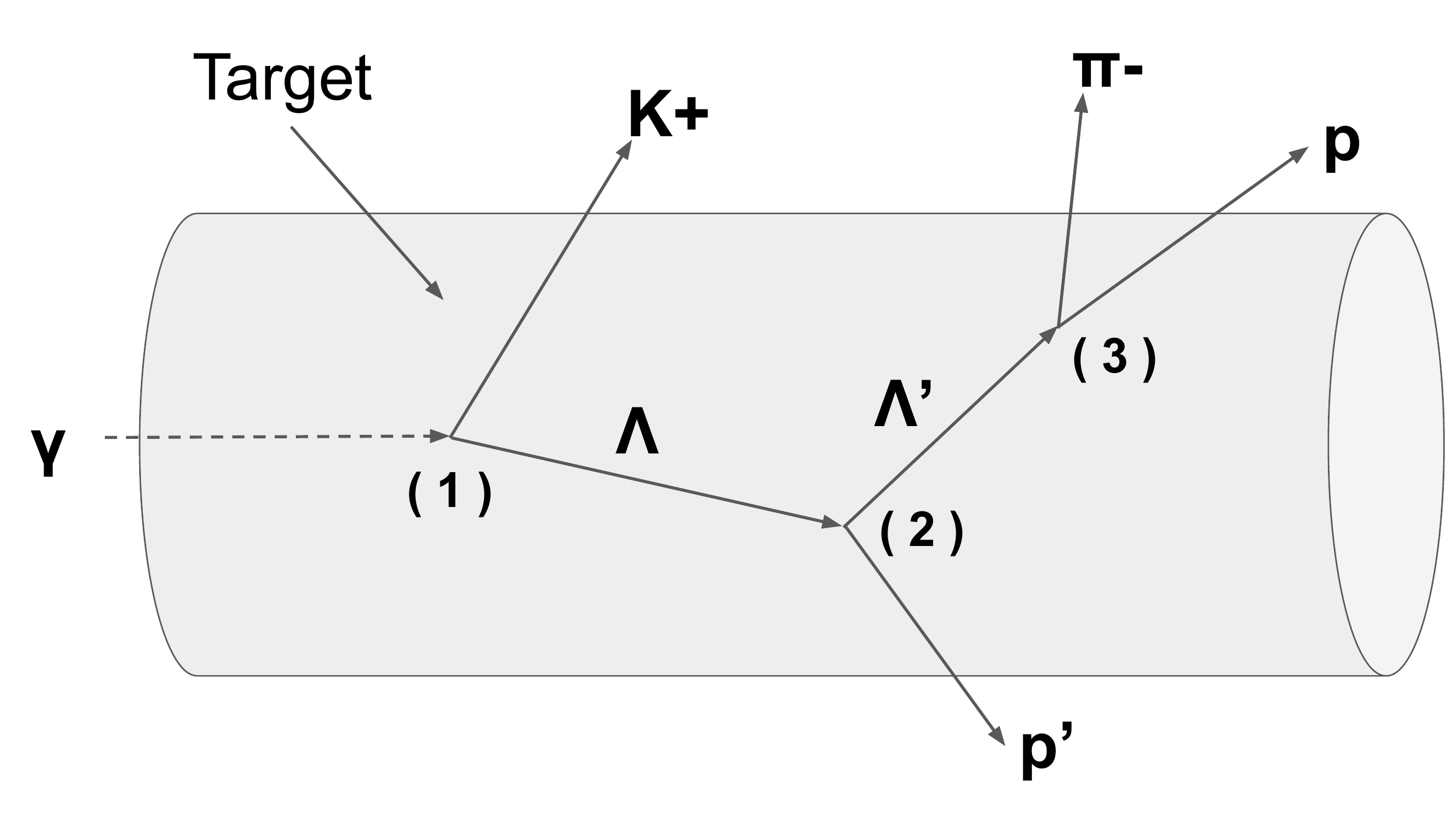}
    \caption{\label{fig:reaction} Pictorial representation of the reaction inside the liquid-hydrogen target. A two-part reaction occurs where the incident $\Lambda$ is created at vertex (1), followed by scattering with a proton at rest in the target at vertex (2), before the $\Lambda$ decays at vertex (3). }
\end{figure}

Observation of $\Lambda p \to \Lambda p$ elastic scattering with the CLAS detector is a two-step process.
The reaction proceeds as follows, which is further illustrated in Fig.~\ref{fig:reaction}:
    \begin{equation}
         \gamma p_{tgt} \rightarrow [K^+] \Lambda ; \Lambda p \rightarrow \Lambda' p' \rightarrow \pi^- p' p .
    \end{equation}
Here, the $\Lambda$ ``beam" is provided from the products of the first reaction.
Detection of the $K^+$, which often decays before reaching the outer part of CLAS, was not required as it was identified using missing mass techniques.
The $\Lambda$ then propagates through the target until it either exits, decays, or scatters with a proton at rest.
The recoil proton, $p'$, was detected directly and the $\Lambda'$ was detected through its decay, $\Lambda' \rightarrow \pi^- p$, which has a branching ratio of 64\%.
The decay proton and $\pi^-$ were directly detected by CLAS, resulting in a fully exclusive measurement.

Final-state particles were detected using standard \textit{g12} techniques. \cite{g12}.
These procedures include timing cuts on the photon and final-state particles, vertex tracing, fiducial region selection, and event trigger efficiency corrections.
The electron beam was bunched into buckets 2-ns apart, which produced the bremsstrahlung photons also in 2-ns bunches.
The final-state particles were filtered using the drift chamber (DC) and time-of-flight scintillator (TOF) for particle identification.
Particle tracks that did not trace back to the target volume were removed.
Fiducial cuts were applied, which filtered out data outside the active region of the DC.
A Monte Carlo (MC) simulation was done to model the CLAS detector in order to measure the reaction acceptance (discussed below).
The simulated events went through the same analysis as the data and included an additional trigger efficiency correction.
An intensive study of the trigger was done \cite{Utsav} and is accounted for in the simulation.

\begin{figure}
		\begin{center}
			\includegraphics[scale = 0.4,keepaspectratio=true]{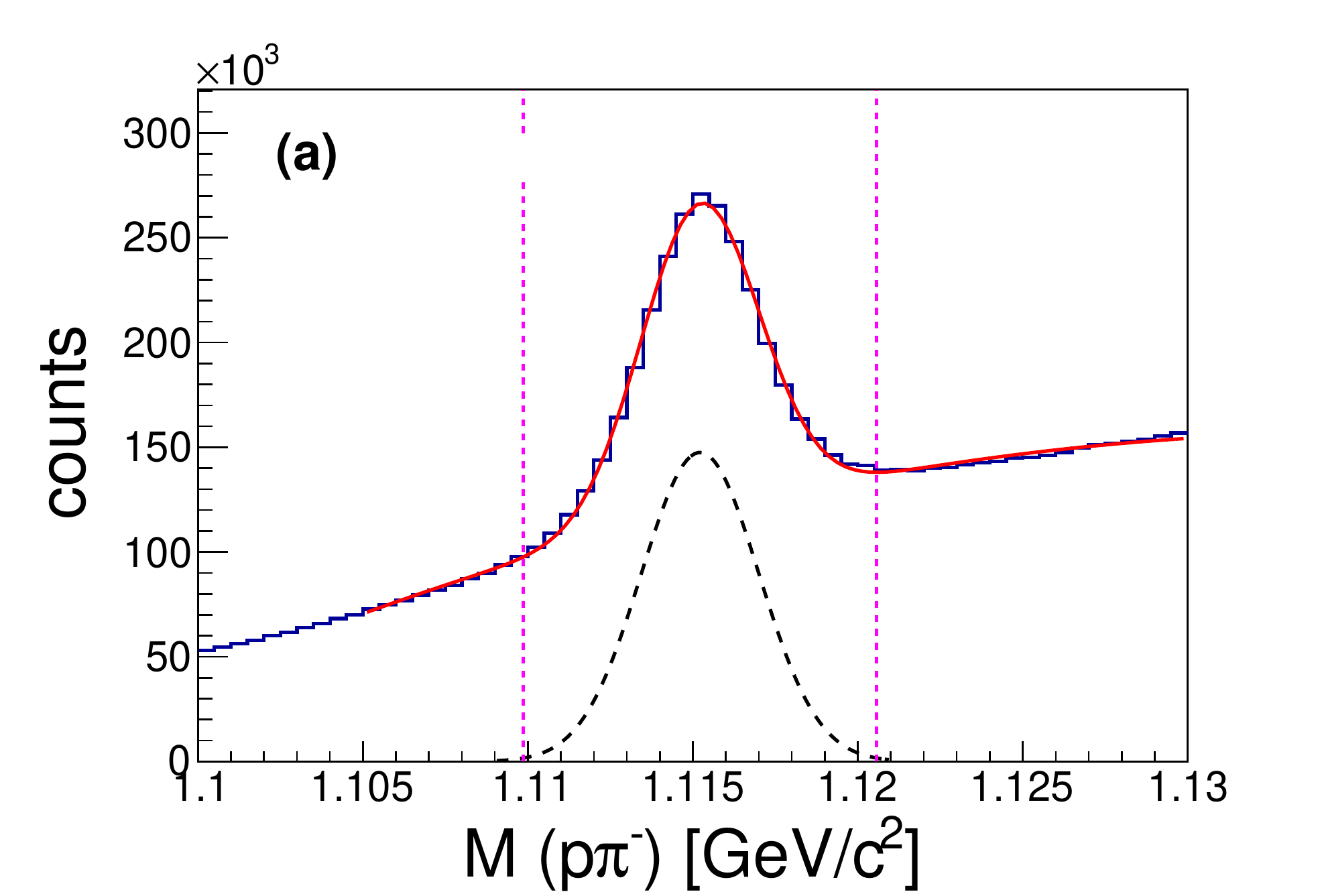}
    		\includegraphics[scale = 0.4,keepaspectratio=true]{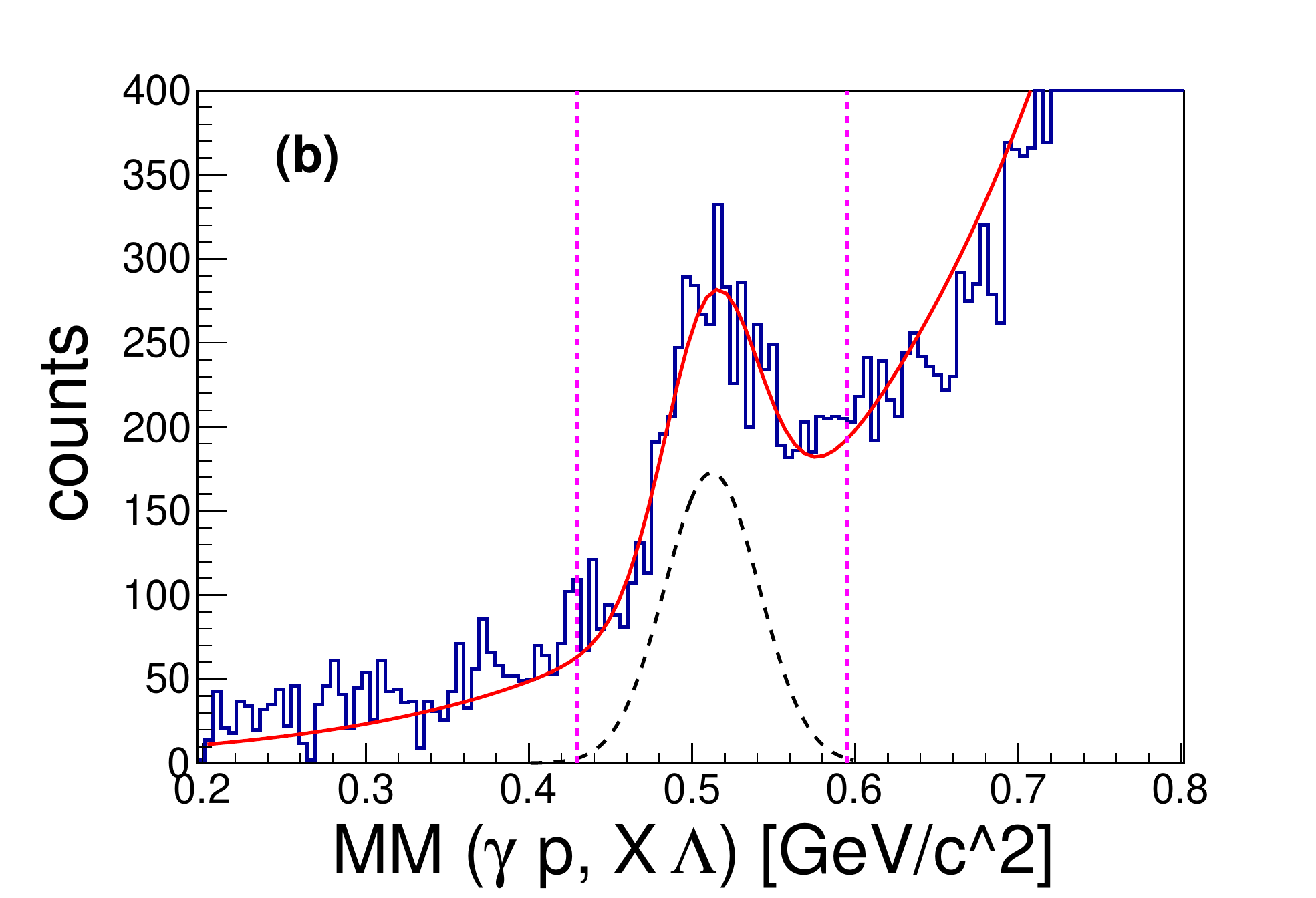}
		\end{center}
		\caption{\label{fig:HlamHmm} Mass spectrum of the detected proton and $\pi^-$ from the decayed $\Lambda '$ (top) and the missing mass spectrum of the initial vertex (bottom). The missing mass spectrum (bottom) was plotted after the cut was made around the $\Lambda '$ mass (top). The total fit, peak plus background, is shown by the solid line. The peak only, a Gaussian, is shown by the dashed line. The vertical lines frame the data that pass through to the final analysis.}
	\end{figure}

The reaction specific analysis required the reaction $\gamma p \rightarrow K^+ \Lambda$ to be isolated.
The scattering $\Lambda'$ was identified from the combined momenta of its decay products, $p_{\pi^-} + p_p$.
These four-momenta produced a mass spectrum, shown in Fig.~\ref{fig:HlamHmm}a.
The peak at 1.115 GeV/c$^2$ corresponds to the scattered $\Lambda '$.
The peak was fit to a Gaussian function, shown by the dashed line.
The data were selected at $\pm 3 \sigma$ for further analysis (see Fig.~\ref{fig:HlamHmm}b).
From the scattered $\Lambda '$ and the other detected proton, the $K^+$ can be identified through the missing four-momentum:
    \begin{eqnarray}\label{eq:mm}
    	p_{X} = p_{\gamma} + p_{tgt} - (p_{\Lambda '} + p_{p'} - p_{tgt}) ,
    \end{eqnarray}
where $p_X$ is the four-momentum of the missing mass distribution,  $p_{\Lambda '}$ is for the recoil $\Lambda$, $p_{p'}$ is for the recoil proton, and $p_{tgt}$ is for target proton.
There are two $p_{tgt}$ terms above, which come from the two target protons at vertex 1 and 2 in Fig.~\ref{fig:reaction}.
This four-momentum gives the missing mass (MM) spectrum shown in Fig.~\ref{fig:HlamHmm}b.
There is a prominent peak at the mass of the $K^+$, 493.7 MeV/c$^2$, which isolates the first vertex of the two-step process leading to the $\Lambda p \rightarrow \Lambda' p'$ elastic scattering.
The peak at the $K^+$ mass was fit to a Gaussian function and a selection was made at $\pm 3 \sigma$.
The background that exists to the right of the $K^+$ mass is due to additional particles produced in the reaction process.
For example, some events may include extra particles such as $\pi^0$ decay of higher-mass $\Lambda^*$ resonances, which were not detected by CLAS.
Those events show up at higher missing mass.

\begin{figure}
        \includegraphics[width=\linewidth]{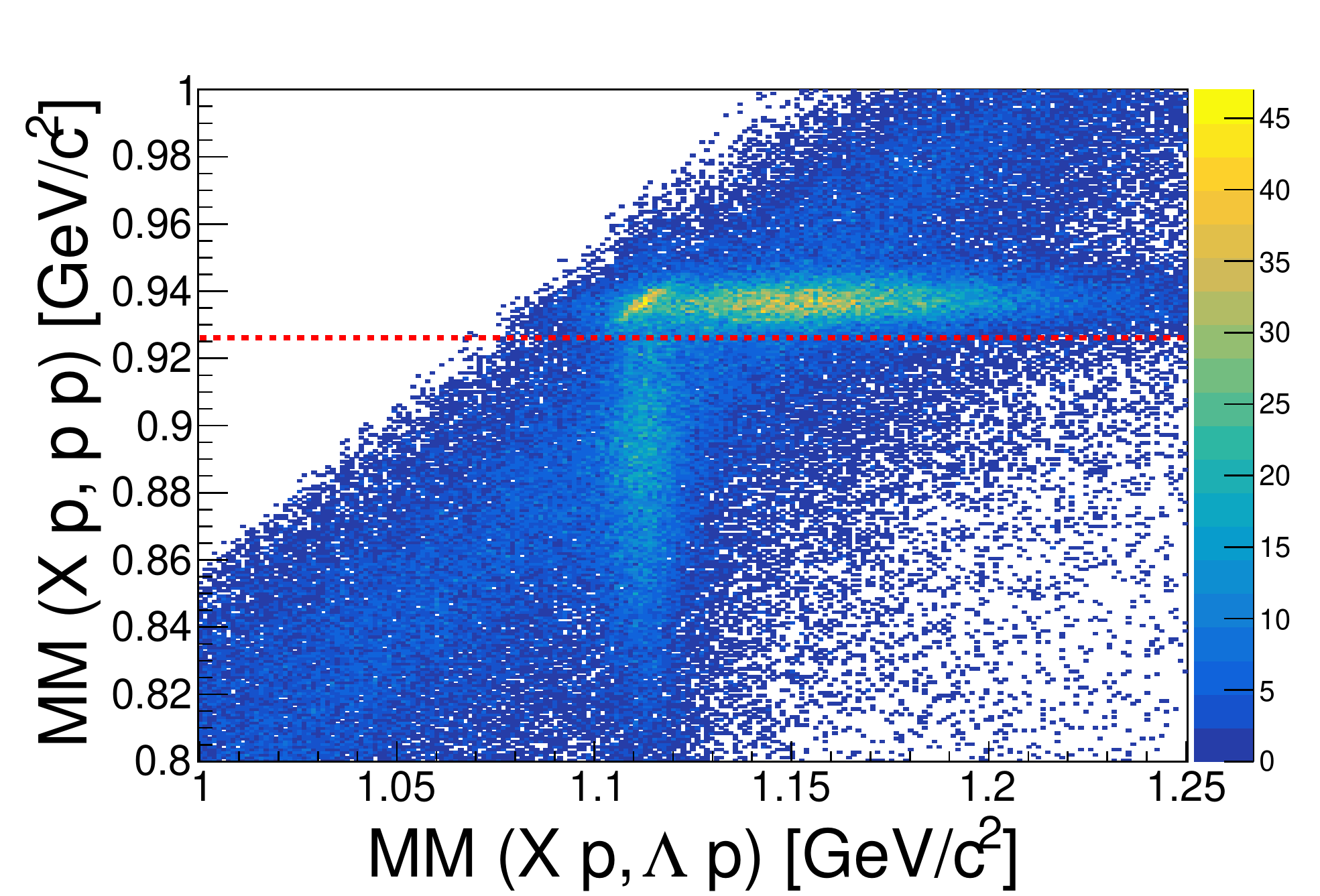}
        \caption{\label{fig:pp} Missing mass spectrum of the secondary vertex $X p \rightarrow \Lambda p$ ($x$-axis) and $Xp \rightarrow pp$ ($y$-axis) where $X$ is the missing mass.}
    \end{figure}

Additional analysis was also required to remove background from the $pp \rightarrow pp$ elastic scattering reaction.
This reaction can happen when the $\Lambda$ decays, followed by an elastic scattering of the decay proton.
This leads to the same final state that can be misidentified as $\Lambda p \to \Lambda' p'$ events.
Kinematic calculations were used to remove these events.
Figure \ref{fig:pp} shows the missing mass distribution of the presumed incident $\Lambda$ on the $x$-axis and that of the presumed proton on the $y$-axis, where X is the missing particle. 
There are prominent bands at the mass of the $\Lambda$ (vertical band) and the mass of the proton (horizontal band).
At the intersection of these bands there is significant overlap.
This region represents $pp$ elastic scattering events that must be removed.
The band to the right of the overlap are $pp$ inelastic scattering events.
Data above the dashed line are rejected, reducing the background along with removing the $pp$ elastic events.
The $\Lambda$ band remains mostly intact.
The same cut is applied to the MC events. 

The $pp$ scattering events were used as a cross check to verify this analysis.
Since many of these events were detected, it was possible to also measure the $pp$ elastic scattering cross section, which is well known.
This method yielded consistent results with the world data for $pp$ scattering.

\begin{figure}
			\includegraphics[scale = 0.4,keepaspectratio=true]{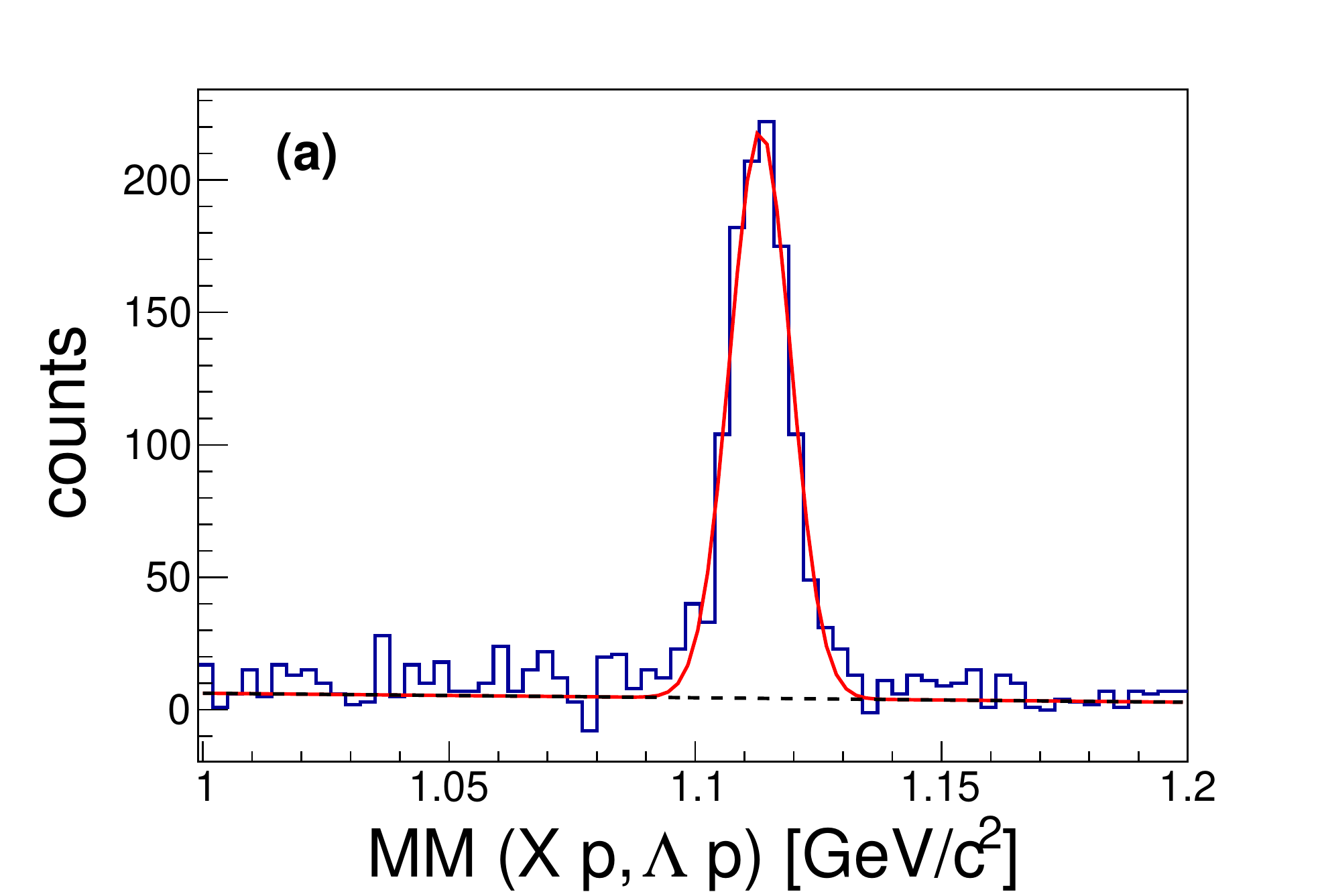}
    		\includegraphics[scale = 0.4,keepaspectratio=true]{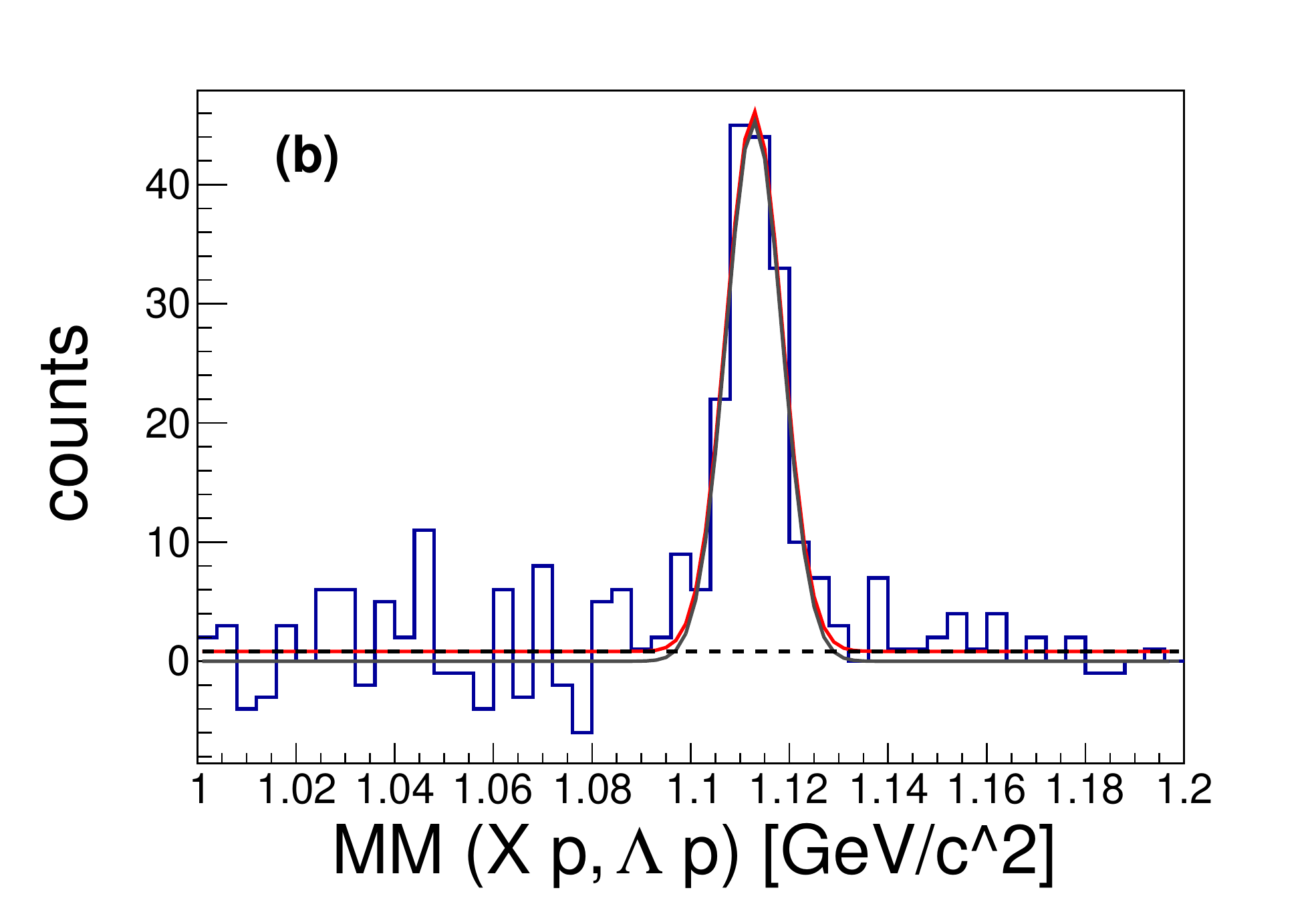}
		\caption{\label{fig:HimYield} Missing mass distribution of the primary vertex integrated over all momentum ranges (top) and binned in the incident $\Lambda$ momentum range 1.3-1.4 GeV after sideband subtraction (bottom).}
	\end{figure}

With the initial reaction $\gamma p \rightarrow K^+ \Lambda$ identified, the incident $\Lambda$ could now be isolated using the missing four momentum:
    \begin{eqnarray}\label{eq:mm}
    	p_{X} = p_{\Lambda '} + p_{p'} - p_{tgt} ,
    \end{eqnarray}
where $p_X$ is for the missing particle, $p_{\Lambda '}$ is for the scattered $\Lambda '$, $p_{p'}$ is for the recoil proton, and $p_{tgt}$ is for the target proton.
The missing mass spectrum of Fig.~\ref{fig:HimYield}a shows a prominent peak at the mass of the $\Lambda$, 1.115 GeV/c$^2$.
This distribution is plotted using events that pass the above selections of both the scattered $\Lambda'$ and the $K^+$ peaks, after subtraction of the background as explained below.
    
The energy dependence of the cross section was determined by binning the missing mass spectrum as a function of the incident $\Lambda$ momentum.
An example bin is shown in Fig.~\ref{fig:HimYield}b.
As part of the analysis, the sideband subtraction technique was used to extract the yield.
This was done by selecting the data to either side of the $\Lambda '$ peak in Fig.~\ref{fig:HlamHmm}a, such that the cut has the same width as that about the peak.
The sideband region should have no scattered $\Lambda$ events, so all the data resulting from sidebands were treated as background and subtracted from the final data.
The sideband subtraction provided a first-order estimate of the background and provided a better signal to noise ratio to extract a yield of the $\Lambda$ peak.
With this method, most of the background was removed from Fig.~\ref{fig:HimYield}, leaving only signal events.
The remaining background in Fig.~\ref{fig:HimYield}b was fit to both a flat line and a second order polynomial.
The flat background was taken as the nominal fit, while the polynomial acted as a check of the systematic uncertainty in the fit.
The signal peak at the mass of the $\Lambda$ was fit to a Gaussian function. The yield was then extracted from the peak fit.

To get the acceptance of the detector, a simulation must be done that models the CLAS detector.
A custom event generator was used to produce $\Lambda p$ elastic scattering events using existing $K^+ \Lambda$ cross sections in order to model a realistic angular dependence \cite{McCracken}. 
The $t$ dependence of the simulation matches that of the data.
Variations in the $t$ parameter were used to study the systematic uncertainty of the acceptance. 
The generated events were passed through a Monte Carlo simulation utilizing the standard GEANT software \cite{GEANT}.
The acceptance of the detector for this two-step reaction ranged from $\sim 0.1 - 2.0$\%. 

Beam flux calculations were more involved than for typical CLAS experiments.
Unlike a photon beam or electron beam that enters the target from one end and is parallel to the beam axis, the $\Lambda$ particles are created throughout the length of the target and have an angular distribution.
The luminosity of the $\Lambda$ beam can be calculated by:
    \begin{equation} \label{eq: Luminosity}
        \mathcal{L}(E_\Lambda) = \frac{N_A \times \rho_T \times l}{M} N_\Lambda(E_\Lambda) ,
    \end{equation}
where $N_A$ is Avogadro's number, $\rho_T$ is the mass density of the target, $l$ is the average path length of the $\Lambda$ beam in the target, $M$ is the molar mass of hydrogen, and $N_\Lambda$ is the number of $\Lambda$ particles in the beam with incident energy $E_\Lambda$.
Unlike the photon beam, for a novel beam like this, the average path length ($l$) and $\Lambda$ flux ($N_{\Lambda}$) cannot be directly measured.
    
To calculate the luminosity of the $\Lambda$ beam, a simulation was made that generated $\Lambda$ particles uniformly throughout the length of the target and within the radius of the photon beam. 
The angular distribution was simulated using known cross sections for the $\gamma p \rightarrow K^+ \Lambda$ vertex \cite{McCracken}.
The simulation also must account for $\Lambda$ particles decaying and exiting the side of the target.
Once the $\Lambda$ particles were generated with their initial properties such as momentum, energy, vertex position, and lab angle, they were propagated through the target.
The probability for particle decay is given by:
	\begin{equation} \label{eq: Probability}
		P(z) = \exp[-\frac{M}{p}\frac{z-z_0}{c\tau}] ,
	\end{equation}
where $P(z)$ is the probability that a $\Lambda$ survives to the point $z$ after being created at $z_0$.
The momentum of the $\Lambda$ is $p/c = M \beta \gamma$ in order to keep everything in the lab frame where the experiment takes place.
The path length was then averaged for each generated particle.

The $\Lambda$ beam flux can be calculated by:
	\begin{equation} \label{eq: LamNumber}
		\sigma = \frac{d\sigma}{d\Omega} (2\pi) (\Delta \cos(\theta)) = \frac{N_\Lambda}{\mathcal{L}_\gamma} ,
	\end{equation}
where $N_\Lambda$ is the number of $\Lambda$, $\mathcal{L}_\gamma$ is the luminosity of the photon beam, and $\theta$ is the center of mass angle of the $K^+$ particle.
The total cross section, $\sigma$, can be calculated from the differential cross section, $d\sigma / d\Omega$, by integrating over the range of $\cos(\theta)$ which is kinematically constrained by the momentum of the particles.
From the simulated path length of a $\Lambda$ for a given momentum, along with the $K^+ \Lambda$ cross sections and the measured photon beam flux, the luminosity of the $\Lambda$ beam was calculated.

\begin{figure}
        \includegraphics[width=\linewidth]{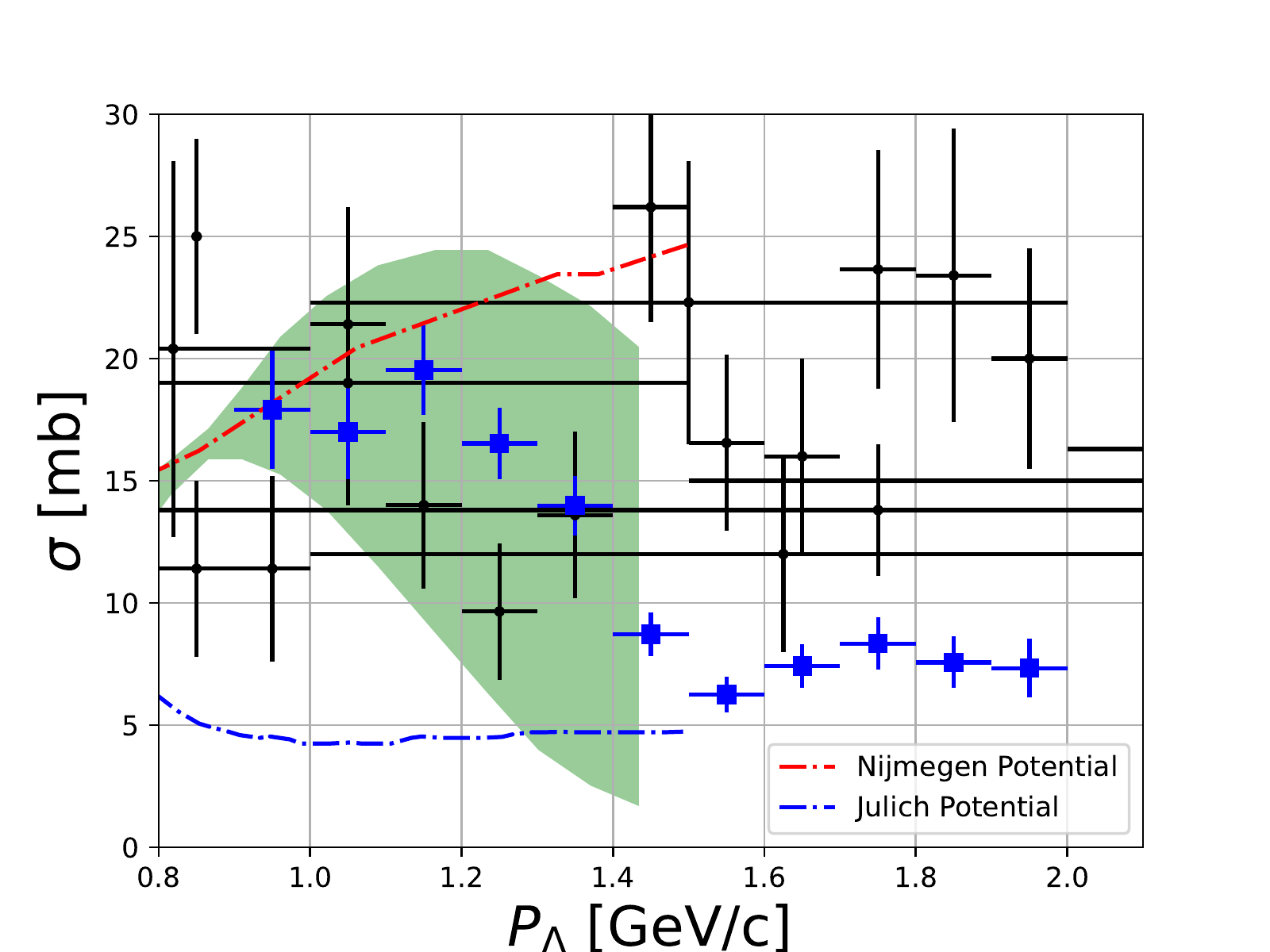}
        \caption{\label{fig:result} Comparison of the total cross section, $\sigma$, with theoretical predictions, extended to a higher $p_{\Lambda}$ using the model of Ref.~\cite{Haidenbauer}. The present results are shown as solid boxes, with previous data as crosses. The green band represents the uncertainty within the chiral EFT model.}
    \end{figure}

Cross sections were calculated for a given momentum bin and integrated over the full angular range as:
    \begin{eqnarray}\label{eq:xs}
        \sigma (p_\Lambda) = \frac{Y (p_\Lambda)}{A (p_\Lambda) \times \mathcal{L} (p_\Lambda) \times \Gamma},
    \end{eqnarray}
where $Y$ is the yield, $A$ is the acceptance for $\Lambda ' p'$, $\mathcal{L}$ is the luminosity of the $\Lambda$ beam, and $\Gamma$ is branching ratio (0.64) \cite{PDG}.
Figure \ref{fig:result} shows the total cross section as a function of the momentum of the incident $\Lambda$ beam.
The data from the present analysis, in solid boxes, are compared to all of the existing world data \cite{PDG,Hauptman}.
The horizontal error bars give the size of the momentum bins.
The vertical error bars represent statistical uncertainties only.

A study of the systematic uncertainties was done at each stage of the analysis.
The largest source comes from the luminosity calculation and the associated $\gamma p \rightarrow K^+ \Lambda$ cross sections.
This systematic uncertainty is estimated at 10.5\%.
The next largest source comes from the choice of $t$-slope dependence in the simulation.
The $t$-parameter dependence was varied in the event generator, which resulted in a systematic uncertainty of 6.7\% for the detector acceptance.
The uncertainty from the beam normalization, described in detail in Ref. \cite{g12}, was estimated at 8.2\%.
Variations in cuts made in the analysis, such as cuts on the $\Lambda '$, $\Lambda$ and $pp$ combined to give an additional uncertainty of 8.4\%.
Other systematic uncertainties include detector-related variations constraints on the detector geometry and target vertex position (2.1\%).
The overall systematic uncertainty for the experiment, added in quadrature, is estimated at 17.3\%.

The theoretical model predictions for $\Lambda p$ elastic scattering, shown by the curves in Fig.~\ref{fig:result}, have been extended to 1.4 GeV for the purpose of this analysis \cite{Haidenbauer-Private}.
Beyond this momentum the theoretical predictions are not reliable.
The green band in Fig.~\ref{fig:result} is a calculation at Next-to-Leading-Order (NLO) from Chiral Effective Field Theory (EFT) \cite{Haidenbauer}.
The derivation of the chiral baryon-baryon potentials for the strangeness sector is done using Weinberg power counting as shown in \cite{Haidenbauer}.
The band represents the range of uncertainty in the calculation.
Fig.~\ref{fig:result} also shows predictions from the well known models of J{\"u}lich \cite{Julich} and Nijmegen \cite{Nijmegen}.
Our measurements do not agree with either of the potentials in this momentum range, but follow close to the center of the EFT calculation.
Furthermore, our results for $p_\Lambda < 1.4$ GeV/c are within the range of previous experiments, but with better precision.
However, our data for $p_\Lambda > 1.4$ GeV/c trend downward, falling below the world data.

A possible explanation for this  could be due to the removal of $pp$ elastic scattering events in the present analysis.
Figure \ref{fig:pp} shows there is strong overlap between $\Lambda p$ and $pp$ elastic scattering events.
Previous bubble chamber experiments did not mention $pp$ elastic scattering as background and, due to the uncertainty in the vertex and energy resolution of bubble chambers, could result in the inclusion of some $pp$ scattering events in those data.
The elastic $pp$ scattering cross sections are several times larger than the elastic $\Lambda p$ cross sections in the momentum range studied here.
If some misidentified $pp$ scattering events remained in the previous experiments, this would increase those cross sections.

We also note that the cross section begins to increase around $p_\Lambda = 1.6$~GeV/c.
This may be due to the opening of an inelastic channel that affects the elastic cross section, likely the reaction $\Lambda p \rightarrow \Lambda (1520) p$.
The threshold for this reaction is at $p_\Lambda = 1.77$ GeV/c, which is a possible explanation for the structure seen in the high momentum range.

To summarize, this experiment was able to improve upon the existing data of $\Lambda p$ elastic scattering in a momentum range of importance to neutron star physics.
We achieve the highest statistical measurement ($<$10\%) for this momentum range.
This is the first experiment to measure the $\Lambda p$ elastic scattering cross sections in this energy range outside of bubble chamber experiments, all of which were done prior to 1980.
The cross sections presented in this paper have higher accuracy, particularly in the higher momentum range, $p_\Lambda > 1.4$~GeV/c, as well as having significantly better precision compared to the existing world data.
These results, along with future three-body reaction data such as the $\Lambda$-deuteron interaction, will help constrain the neutron star EOS. Techniques developed here for secondary scattering of hadrons from photoproduction can also be used for future data analysis.
Measurements at CLAS using this technique on a deuteron target are in progress \cite{Ilieva}.

\hfill \break
The authors acknowledge the staff of the Accelerator and Physics Divisions at the Thomas Jefferson National Accelerator Facility who made this experiment possible. This work was supported in part by the Chilean Comisi\'{o}n Nacional de Investigaci\'{o}n Cient\'{i}fica y Tecnol\'{o}gica (CONICYT), by CONICYT PIA Grant No.\,ACT1413, the Italian Istituto Nazionale di Fisica Nucleare, the French Centre National de la Recherche Scientifique, the French Commissariat \'{a} l’Energie Atomique, the United Kingdom Science and Technology Facilities Council (STFC), the Scottish Universities Physics Alliance (SUPA), the National Research Foundation of Korea, and the U.S. National Science Foundation. The Southeastern Universities Research Association operates the Thomas Jefferson National Accelerator Facility for the United States Department of Energy under Contract No.\,DE-AC05-06OR23177.

\end{document}